\documentclass[12 pt, amsfonts,amsmath, amssymb,color]{article}

\usepackage{amsmath,amssymb,amsfonts,latexsym,float,graphics,epsfig}
\usepackage{subfig}
\usepackage{verbatim}
\usepackage{relsize}
\usepackage{hyperref}

\def\be{\begin{equation}}
\def\ee{\end{equation}}
\def\bea{\begin{eqnarray}}
\def\eea{\end{eqnarray}}

\def\R{\mathbb{R}}

\begin{document}
%<<<<<<<<<<< enumeration of eqns section wise>>>>>>>>>>>>>>>>>>>

\renewcommand\theequation{\arabic{section}.\arabic{equation}}
\catcode`@=11 \@addtoreset{equation}{section}
%<<<<<<<<<<<<<<<<<<<<<<<<<<<<<<<<<>>>>>>>>>>>>>>>>>>>>>>>>>>>>>>>>>
\newtheorem{axiom}{Definition}[section]
\newtheorem{theorem}{Theorem}[section]
\newtheorem{axiom2}{Example}[section]
\newtheorem{lem}{Lemma}[section]
\newtheorem{prop}{Proposition}[section]
\newtheorem{cor}{Corollary}[section]

\newcommand{\ben}{\begin{equation*}}
\newcommand{\een}{\end{equation*}}
\title{\bf Thermodynamic geometry for charged Gauss-Bonnet black holes in AdS spacetimes}
\author{
\bf Aritra Ghosh\footnote{E-mail: ag34@iitbbs.ac.in} \hspace{0.5mm} and Chandrasekhar Bhamidipati\footnote{E-mail: chandrasekhar@iitbbs.ac.in} \\
~~~~~\\
 School of Basic Sciences, Indian Institute of Technology Bhubaneswar,\\   Jatni, Khurda, Odisha, 752050, India\\
}

\date{ }

\maketitle

\begin{abstract}
In this paper, we study the thermodynamic geometry of charged Gauss-Bonnet black holes (and Reissner-Nordstr\"{o}m black holes, for the sake of comparison) in AdS:  in both \((T,V)\)- and \((S,P)\)-planes. The thermodynamic phase space is known to have an underlying contact and metric structure; Ruppeiner geometry then naturally arises in this framework. Sign of Ruppeiner curvature can be used to probe the nature of interactions between the black hole microstructures. It is found that there are both attraction and repulsion dominated regions which are in general determined by the electric charge, Gauss-Bonnet coupling and horizon radius of the black hole. The results are physically explained by considering that these black hole systems consist of charged as well as neutral microstructures much like a binary mixture of fluids.
\end{abstract}

% \maketitle
\section{Introduction}
Since the work of Bekenstein \cite{Bekenstein:1973ur,Bekenstein:1974ax}, Hawking \cite{Hawking:1974sw,Hawking:1976de} and related developments \cite{Bardeen:1973gs,Gibbons:1976ue} black hole thermodynamics has been an exciting and ongoing field of active research with interesting results. The entropy of a black hole has been identified with the area of its event horizon while the temperature with the surface gravity of the black hole. The first law of black hole thermodynamics has been stated as,
\begin{equation}
\label{first law non extended}
dM = TdS + \Phi dQ,
\end{equation}
where \(\Phi\) corresponds to the electric potential and \(Q\) is the electric charge of the black hole. The mass \(M\) of the black hole was identified the internal energy \(U\) in the conventional thermodynamic treatments. The first law [eqn(\ref{first law non extended})] is by no means trivial. It means that the ``hairs" of a black hole vary in a manner analogous to the first law of standard thermodynamics, a fact that is quite fascinating in itself. Since its inception, there has been a flurry of research activity on black holes and their thermodynamics particularly in Anti de Sitter (AdS) spacetimes~\cite{Hawking:1982dh}. Due to the lack of \(P\) and \(V\) terms in black hole thermodynamics, phase transitions were studied on the \((Q,T)\)-phase plane as compared to \((P,T)\)-phase diagrams in standard thermodynamic treatments~\cite{Chamblin,Chamblin:1999hg}. However, the recent revival of pressure \(P\) and volume \(V\) in black hole thermodynamics, where the cosmological constant \(\Lambda\) is taken to be dynamical and treated as thermodynamic pressure, has led to interesting developments \cite{Henneaux:1984ji}-\cite{Kubiznak:2016qmn}. It has been shown that black holes in AdS admit equations of state similar to that of a van der Waals fluid (see for example \cite{Kubiznak:2012wp, Gunasekaran:2012dq}). The first law of black hole thermodynamics has thus been restated as,
\begin{equation}
\label{first law extended}
dM = VdP + TdS + \Phi dQ,
\end{equation} where in this ``extended" thermodynamic phase space, the mass \(M\) of the black hole is identified with its enthalpy, \(H(S,P,Q) = U + PV\).\\

%\smallskip \noindent
\noindent
In 1979, Ruppeiner \cite{Ruppeiner} introduced a metric on the space of thermodynamic equilibrium states defined as the negative Hessian of the entropy. Ruppeiner geometry \cite{Ruppeiner:1995zz}-\cite{Ruppeiner3} is widely used in thermodynamic fluctuation theory and particularly in black hole thermodynamics since the singularities in the Ruppeiner curvature indicate critical points~\cite{Ruppeiner:1995zz}. The thermodynamic length between two states is calculated using the Ruppeiner line element defined as,
\begin{equation}\label{SRuppeiner}
  ds_R^2 = - \frac{\partial^2S}{\partial x^i \partial x^j}dx^idx^j,
\end{equation}
where \(x^i\) are independent thermodynamic variables and \(i,j \in \{1,2,...,n\}\). In the context of fluctuation theory, the farther apart two points are on the space of thermodynamic equilibrium states, the less probable are fluctuations between such states. Moreover, in standard thermodynamics it has been found that the Ruppeiner metric is flat for systems where the molecules are non-interacting. A non-zero Ruppeiner curvature indicates interactions between the underlying molecules forming the thermodynamic system. A negative curvature in the Ruppeiner metric indicates an overall attractive interaction whereas a positive curvature indicates repulsive interactions between the molecules. It has been suggested that black holes can be associated with microstructures (see \cite{Wei:2015iwa,AR, Mann2019,Wei:2019yvs,SAdS} and references therein) analogous to molecules constituting a macroscopic system in standard thermodynamics. Such microstructures may be characterized by repulsive as well as attractive interactions between them if the corresponding Ruppeiner metric is curved. Therefore, Ruppeiner geometry can be used to understand the nature of interactions between microstructures, which to the best of our knowledge, was first used in the context of BTZ black holes~\cite{Banados:1992wn} in three dimensions in~\cite{Cai:1998ep}, to gain a statistical understanding of the underlying degrees of freedom. Furthermore in~\cite{Shen:2005nu}, Ruppeiner geometry for Reissner-Nordstr\"{o}m, Kerr and Reissner-Nordstr\"{o}m-AdS black holes was studied in the non-extended phase space, using eqn (\ref{SRuppeiner}), with internal energy and electric potential (or angular velocity for Kerr black hole case) as the fluctuation variables. It was also shown that the divergence of the scalar curvature is consistent with the Davies' phase transition point~\cite{Davies:1978mf}. Following these results, this method has now been generalized to understand several systems~\cite{Dehyadegari,Moumni,Deng,Sheykhi,Miao,Miao2,Miao3,Li,Chen,Du,Xuz}. In \cite{Wei:2015iwa,Mann2019,Wei:2019yvs}, Ruppeiner geometry for Reissner-Nordstr\"{o}m-AdS black holes\footnote{See~\cite{Sahay:2010tx,Ruppeiner:2011gm,Sahay:2016kex,Chaturvedi:2017vgq} where Ruppeiner metric for black hole systems in other parameterizations were considered.} has been explored on the \((T,V)\)-plane and interesting results on the microscopic description are presented. The Schwarzschild black hole in AdS has been studied quite recently \cite{SAdS} using Ruppeiner geometry in an enthalpy representation, i.e. on the \((S,P)\)-plane and it has been shown that attractive interactions dominate between the black holes microstructures.\\

%\smallskip \noindent
\noindent
It is now well known that the thermodynamic phase space assumes a contact structure \cite{hermann}-\cite{mrugala1996}. In \cite{contactBH}, we have studied the extended phase space thermodynamics of black holes in AdS in the contact geometry framework. The thermodynamic transformations are expressed as flows on Legendre submanifolds\footnote{Alternatively known as equilibrium submanifolds or spaces of thermodynamic equilibrium states.} of the thermodynamic phase space. These submanifolds are the spaces of equilibrium states where each point corresponds to a thermodynamic equilibrium state. Mrugula and others \cite{therm} - \cite{mrugala1996} have shown that a Riemannian metric that is in a sense compatible with the contact structure can be defined in the thermodynamic phase space. The Ruppeiner metric is equivalent to the metric introduced by Mrugula when restricted to Legendre submanifolds. There are some subtle issues which require Legendre invariance of thermodynamic metrics (see for example \cite{Quevedo:2006xk,Quevedo:2007mj,Zhang:2015ova,Hendi:2015xya}) and connections between metrics derived from other potentials. We shall however not address such issues here.\\

%\smallskip \noindent
\noindent
\textbf{Motivation and plan:} The primary motivation of this paper is to study Ruppeiner geometry and hence probe the nature of interactions between the microstructures in charged Gauss-Bonnet black holes in AdS within the framework of extended phase space thermodynamics. For the sake of completeness we shall recall the compatible metric structure of the thermodynamic phase space which reduces to the Ruppeiner metric on the spaces of thermodynamic equilibrium states or the so called Legendre submanifolds. We then study the Ruppeiner geometry and obtain the corresponding curvature scalars on both the \((S,P)\)- and \((T,V)\)- planes using enthalpy and Helmholtz free energy representations respectively for the charged Gauss-Bonnet black holes in AdS\(_5\)(see~\cite{Wei:2012ui,Wei:2019ctz} for recent developments). We comment on the nature of interactions and compare the results with those for Reissner-Nordstr\"{o}m black holes in AdS\(_4\). \\
%\smallskip \noindent

\noindent
The \underline{paper is organized} as follows. In the following section, we quickly recall some elements of contact geometry as well as the compatible metric structures on contact manifolds. We then very briefly review black hole thermodynamics in the extended phase space from a contact geometry perspective in the same section. Ruppeiner geometry for Reissner-Nordstr\"{o}m-AdS black holes is presented in section (3) while section (4) is dedicated to the study of Ruppeiner geometry for charged Gauss-Bonnet-AdS black holes. The physical interpretations of the results are discussed. Finally, we end the paper with remarks in section (5). In Appendix-A we collect few details of contact Hamiltonian dynamics and in Appendix-B, we show how the thermodynamics of GB-AdS black holes emerges from the high temperature ideal gas limit via suitable deformations.

\section{The geometry of thermodynamics}
In this section, we quickly review some aspects of contact geometry, metric structures and hence Ruppeiner geometry in the context of black hole thermodynamics. This will set up the background for the rest of the paper to study metric structures in extended phase space thermodynamics of black holes in AdS.

\subsection{Contact geometry}
Contact geometry \cite{Geiges, Arnold} is suitable for description of simple dissipative systems in mechanics \cite{CM1, CM2}. It has also been applied to thermodynamics \cite{RT1} - \cite{RT2} as well as to statistical mechanics \cite{SM}. We shall briefly recall the elements of contact geometry first and then look at compatible metric structures on a contact manifold. Recall that a contact manifold is the pair \((\mathcal{M},\eta)\) where \(\mathcal{M}\) is a smooth manifold of dimension \((2n+1)\) and \(\eta\) is a 1-form that satisfies the condition of complete non-integrability, \begin{equation}\label{nonint}
\eta \wedge (d\eta)^n \neq 0.
\end{equation}
This condition implies that one can write a Whitney sum decomposition of the tangent bundle of \(\mathcal{M}\) as,
\begin{equation}\label{splittingTM}
  T\mathcal{M} = \mathcal{V} \oplus \mathcal{H},
\end{equation}
where, the distributions $\mathcal{V} = ker(d\eta)$ and $\mathcal{H} = ker(\eta)$ are both regular and are respectively known as vertical and horizontal distributions. It is clear that the dimension of \(ker(d\eta)\) is one. In fact, \(d\eta\) is non-degenerate when restricted to \(ker(\eta)\). We may identify \(\eta \wedge (d\eta)^n\) with a standard volume form on the \(2n+1\) dimensional \(\mathcal{M}\).\\

%\smallskip \noindent
\noindent
On \(\mathcal{M}\) there exists a unique and global vector field \(\xi\) known as the Reeb vector field determined by the conditions,
\begin{equation}\label{reeb}
  \eta(\xi) = 1, \hspace{3mm} d\eta(\xi,.) = 0.
\end{equation}
The Reeb vector field results in the natural splitting of \(T\mathcal{M}\) since \(\eta(\xi) = 1\) means that the vector field \(\xi\) generates a 1-dimensional distribution that is complimentary to \(2n\)-dimensional horizontral distribution \(\mathcal{H} = ker(\eta)\). There exists another fundamental object \(\phi\), a \((1,1)\) tensor field associated with \(\eta\) such that, \(\mathcal{V} = ker(\phi)\) and \(\mathcal{H} = Im(\phi)\) so that eqn (\ref{splittingTM}) can be re-written as,
\begin{equation}\label{splittingTM1}
  T\mathcal{M} = ker(\phi) \oplus Im(\phi).
\end{equation}
Moreoever, if the tensor field \(\phi\) satisfies the following identity,
\begin{equation}\label{phi}
  \phi^2 = - I + \eta \otimes \xi,
\end{equation} where \(I\) is the identity operator then \(\mathcal{M}\) is said to have a \((\eta, \xi, \phi)\) structure. Eqn (\ref{phi}) implies that given a vector \(X \in T\mathcal{M}\), it is possible to write a decomposition in vertical and horizontal components,
\begin{equation}\label{verhorX}
  X =  \eta(X)\xi - \phi^2 X,
\end{equation}
where \(X_v = \eta(X)\xi\) and \(X_h = - \phi^2 X\) are respectively the vertical and horizontal components of \(X\).\\

\noindent
In local (Darboux) coordinates \((s,q^i,p_i)\) on \(\mathcal{M}\), the expressions for \(\eta\) and \(\xi\) are,
\begin{equation}\label{etaxicoordinateexpression}
  \eta = ds - p_idq^i; \hspace{3mm} \xi = \frac{\partial}{\partial s}.
\end{equation}
It is easy to check using eqns (\ref{etaxicoordinateexpression}) that the conditions in eqn (\ref{reeb}) are satisfied.

\subsection{Legendre submanifolds}
We now define a very special class of submanifolds of a contact manifold. These are the Legendre submanifolds analogous to the Lagrangian submanifolds of symplectic manifolds. Naively, they are described as maximal dimensional solutions of the equation \(\eta=0\).\\

\noindent
Consider a submanifold \(L \subset \mathcal{M}\) of the thermodynamic phase space. Then, if \(\eta|_L=0\) then such a submanifold is called a isotropic submanifold. This means that coordinates on an isotropic submanifold cannot include a conjugate pair of variables. If however, \(L\) is a maximal dimensional isotropic submanifold then it is called a Legendre submanifold. It follows that the maximal dimension of \(L\) is \(n\) and hence, all Legendre submanifolds have dimension equal to \(n\). The local form of such a submanifold is expressed as,
\begin{equation}\label{legendrelocal}
   p_i = \frac{\partial F}{\partial q^i}, \hspace{3mm} q^j = -\frac{\partial F}{\partial p_j},  \hspace{3mm} s = F - p_j \frac{\partial F}{\partial p_j}.
\end{equation} Here \(F=F(q^i,p_j)\) is known as the generator of the Legendre submanifold and where \(I \cup J\) is a disjoint partition of the set of indices \(\{1,2,....,n\}\) with \(i \in I, j \in J\).

\subsection{Metric structure on a contact manifold}
We shall now describe compatible metric structures \cite{mrugala1990, mrugala1996, bravetti2015} on a contact manifold. If a contact manifold \(\mathcal{M}\) with a \((\eta, \xi, \phi)\) structure can be associated with a metric \(G\) such that, \begin{equation}\label{G1}
  G(\phi X_1, \phi X_2) = G(X_1,X_2) - \eta(X_1)\eta(X_2),
\end{equation}
then \(G\) is called an compatible Riemannian metric and \((\mathcal{M}, \eta, \xi, \phi)\) a contact metric manifold. It was shown \cite{A}-\cite{C} that given a contact manifold \((\mathcal{M},\eta)\), it is always possible to associate a metric structure. The metric \(G\) is as a bilinear, symmetric and non-degenerate form on \(\mathcal{M}\). It is taken to be of the form ,
\begin{equation}\label{metriclocal}
  G = \eta^2 - dp_idq^i.
\end{equation}
One can check that the form of \(G\) taken in eqn (\ref{metriclocal}) is bilinear, symmetric as well as non-degenerate. Now consider a particular Legendre submanifold \(L\) with \(s = F(q^i)\) with \(i=\{1,...,n\}\) so that from eqns (\ref{legendrelocal}), the local form of \(L\) is,
\begin{equation}\label{thermodynamiclocal}
  s = F(q^i); \hspace{3mm} p_i = \frac{\partial F(q^i)}{\partial q^i}.
\end{equation}
Since \(\eta|_L=0\) it immediately follows that restricting to \(L\), the metric \(G\) takes the local form,
\begin{equation}\label{GLlocal}
  G|_L = -dp_idq^i|_L = -\frac{\partial^2 F}{\partial q^i \partial q^{i'}}dq^idq^{i'}; \hspace{
  5mm} i,i' \in \{1,2,....,n\}.
\end{equation}
The metric tensor is therefore the negative of the Hessian matrix of \(F\) which is generating function of the Legendre submanifold \(L\). In the thermodynamic case, the generator \(F\) is an appropriate thermodynamic potential and hence on a Legendre submanifold or space of thermodynamic equilibrium states, the metric is the negative Hessian of the thermodynamic potential. This means that \(G|_L\) is a generalized Ruppeiner metric \cite{Ruppeiner} on the space of thermodynamic equilibrium states defined as the negative Hessian of the potential or its conformal equivalent Weinhold metric \cite{Weinhold}.\\

\noindent
It is straightforward to generalize eqn (\ref{GLlocal}) to an arbitrary Legendre submanifold \(\tilde{L}\) with a generating function of the form \(\tilde{F}=\tilde{F}(q^i,p_j)\). From eqns (\ref{legendrelocal}) a simple calculation shows that when restricted to \(\tilde{L}\) the metric takes the following form,
\begin{equation}
  g = -dp_idq^i|_{\tilde{L}} = \frac{\partial^2 \tilde{F}}{\partial p_j \partial p_{j'}}dp_jdp_{j'} - \frac{\partial^2 \tilde{F}}{\partial q^i \partial q^{i'}}dq^idq^{i'},
\end{equation}
with \(i,i' \in I\) and \(j,j' \in J\) where \(I\) and \(J\) are a disjoint partition of indices with \(I \cup J = \{1,2,....,n\}\). This generalizes the metric of eqn (\ref{GLlocal}) to an arbitrary potential \(\tilde{F}=\tilde{F}(q^i,p_j)\). However, we remark that in reversible thermodynamics, only Legendre submanifolds of the local form given by eqns (\ref{thermodynamiclocal}) appear and hence Ruppeiner geometry is a standard framework for the study of thermodynamic metric structures.

\subsection{Thermodynamic metrics for black holes}\label{thermodynamicsBH}
We now recall the first law of black hole thermodynamics in the extended phase space,
\begin{equation}\label{firstlawH}
  dH - TdS - VdP - \Phi dQ = 0,
\end{equation} where symbols have their usual meanings. A direct comparison of equations (\ref{etaxicoordinateexpression}) and (\ref{firstlawH}) allows for the identification that the thermodynamic coordinates \(\{U,T,S,P,V,\Phi,Q\}\) are coordinates on a contact manifold \((\mathcal{M},\eta)\). The Darboux coordinates in this case are, \(s=H, q^1=S, q^2=P, q^3 = Q, p_1 = T, p_2 =V, p_3 = \Phi\). This seven dimensional contact manifold is the thermodynamic phase space. Alternatively, one could have written eqn (\ref{firstlawH}) as,
\begin{equation}\label{firstlawS}
  dS - \beta dH + \beta V dP + \beta \Phi dQ = 0,
\end{equation} where we may identify coordinates on \(\mathcal{M}\) as, \(s = S, q^1=H, q^2=P, q^3 = Q, p_1 = \beta, p_2 =-\beta V, p_3 = - \beta \Phi\) where \(\beta = 1/T\) is the inverse temperature factor. Therefore, the local coordinates \((s,q^i,p_i)\) need not have a unique identification.\\

 \noindent
From eqn (\ref{firstlawH}) we straightforwardly obtain,
\begin{equation}\label{legendresubmanifoldH}
  s = H; \hspace{3mm} T = \frac{\partial H}{\partial S}; \hspace{3mm}  V = \frac{\partial H}{\partial P}; \hspace{3mm}  \Phi = \frac{\partial H}{\partial Q}.
\end{equation}
These equations locally describe the Legendre submanifold \(L\) representing the black hole. The generator of this submanifold \(L\) turns out to be the enthalpy \(H\). The space of equilibrium states \(L\) is a Legendre submanifold of the entire thermodynamic phase space \(\mathcal{M}\) with a smooth inclusion map,
$$ f : L \rightarrow \mathcal{M}. $$
\noindent
As remarked in the previous subsection, in thermodynamics, the Legendre submanifolds that appear are of the local form given by eqns (\ref{thermodynamiclocal}) wherein the thermodynamic potential is identified as being a function of the so called "coordinates" \(\{q^i\}\) only and this motivates a Hamilton-Jacobi theory for black hole thermodynamics \cite{Rajeev1,contactBH}. Moreover, as a consequence the thermodynamic metric takes the form of eqn (\ref{GLlocal}). It is straightforward to show that the Ruppeiner line element defined in eqn (\ref{SRuppeiner}) on the \((S,P)\)-plane i.e., in an enthalpy representation, takes the form,
\begin{equation}\label{RuppeinerSP}
  ds_R^2 = \frac{1}{C_P}dS^2 + \frac{2}{T}\bigg(\frac{\partial T}{\partial P}\bigg)_SdSdP,
\end{equation}
where, the \(dP^2\) term vanishes since \((\partial^2H/\partial P^2)_S = (\partial V/\partial P)_S = 0\) for a static black holes. However, if one considers a Legendre submanifold with the independent thermodynamic coordinates as \(T\) and \(V\), then the corresponding Ruppeiner line element [eqn (\ref{SRuppeiner})] in this Helmholtz free energy representation has the form,
\begin{equation}\label{RuppeinerTV}
  ds_R^2 = \frac{1}{T}\bigg(\frac{\partial P}{\partial V}\bigg)_TdV^2 + \frac{2}{T}\bigg(\frac{\partial P}{\partial T}\bigg)_VdTdV,
\end{equation}
where the \(dT^2\) term vanishes since, \((\partial^2 F/\partial T^2)_V = -(\partial S/\partial T)_V = 0\) for static black holes. In the following sections, we shall use eqns (\ref{RuppeinerSP}) and (\ref{RuppeinerTV}) to obtain the thermodynamic line elements for black holes. Let us note that, a metric in $(T,V)$ plane was proposed to study microstructures (assuming $C_V$ for static black holes to be nonzero) of Reissner-Nordstr\"{o}m black holes in~\cite{Mann2019}, and for neutral Gauss-Bonnet black holes in AdS in~\cite{Wei:2019ctz}. The metric in eqn (\ref{RuppeinerTV}) above is however different from the one in~\cite{Mann2019,Wei:2019ctz}, as it is valid even for $C_V=0$. More comments are made in the sections below. Further, the  metrics in eqns (\ref{RuppeinerSP}) and (\ref{RuppeinerTV}) were also proposed recently in~\cite{SAdS} to study the thermodynamic geometry of Schwarzschild black holes in AdS.

%%%%%%%%%%%%%%%%%%%%%%%%%%%%%%%%%%%%%
\section{Reissner-Nordstr\"{o}m black holes in AdS}
%%%%%%%%%%%%%%%%%%%%%%%%%%%%%%%%%%%%
We now consider Reissner-Nordstr\"{o}m (RN-AdS) black holes in AdS\(_4\) with fixed charge. The mass \(M\) of the black hole, equated to its enthalpy \(H\) is given as,
\begin{equation}\label{HRN}
H(S,P)= M = \frac {1} {6\sqrt{\pi}} S^{-\frac{1}{2}}
\left(8 P S^2 + 3S+3\pi q^2 \right).
\end{equation}
The thermodynamic equation of state for the RN-AdS black hole when expressed in terms of the specific volume \(v=2r_+\) is given as,
\begin{equation}\label{RNeqnofstatev}
P=\frac{T}{v}-\frac{1}{2\pi v^2}+\frac{2q^2}{\pi v^4}\,.
\end{equation}
This equation of state can be viewed as a deformed equation of state of the non-interacting ideal gas with deformation terms in the pressure~\cite{contactBH} arising due to interactions, which in this case are the last two terms in eqn (\ref{RNeqnofstatev}). The first interaction term carrying the negative sign is of the van der Waals type and is clearly due to attractive interactions while the second term carrying a positive sign is of the repulsive type. Therefore, there is a competition between the attractive and repulsive interactions between microstructures in case of a RN-AdS black hole. This shall lead to, as we shall see attraction or repulsion dominated cases depending which interaction dominates over the other.\\

%This becomes more transparent if we re-write the equation of state as:
%\begin{equation}
 % P = P^0_{RN} + P^i_{RN}
%\end{equation}
%where \(P^0_{RN} = T/v\) is the pressure of the ideal gas while the terms due to interactions are given by:
%\begin{equation}\label{RNP_i}
 % P^i_{RN} = -\frac{1}{2\pi v^2} + \frac{2q^2}{\pi v^4}
%\end{equation}
%The first term in eqn (\ref{RNP_i}) is a Van der Waals fluid-like term and is due to attractive interactions. The second term is with a positive sign and naturally arises due to repulsive interactions. Therefore, there is a competition between the attractive and repulsive interactions between microstructures in case of a RN-AdS black hole. This shall lead to, as we shall see attraction or repulsion dominated cases depending which interaction dominates over the other.

\begin{figure}[h]
%	 \begin{wrapfigure}{l}{0.3\textwidth}
	\begin{center}
		\centering
		\includegraphics[width=4.6in]{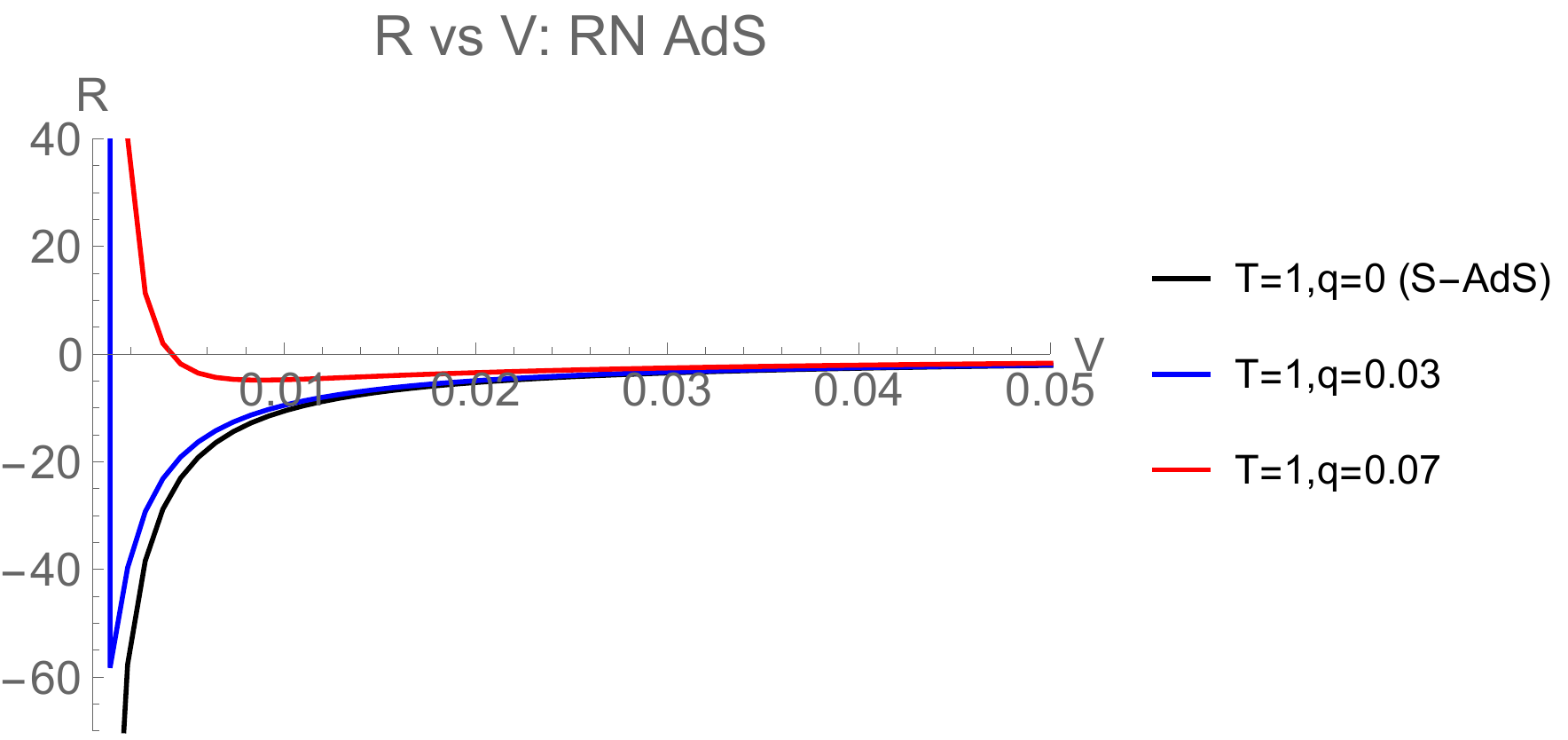}  		
		\caption{Curvature of Ruppeiner metric on \((T,V)\)-plane for RN-AdS black holes as a function of \(V\) for fixed \(T\). The case with \(q=0\) corresponds to the Schwarzschild-AdS black holes.}   \label{R_V_RNAdS}		
	\end{center}
%	\end{wrapfigure}
\end{figure}

 \noindent
On the \((T,V)\)-plane, the Ricci scalar associated with the Ruppeiner metric is computed to be,
\begin{equation}
R_{TV} =\frac{\big(4 \sqrt[3]{6} \pi ^{2/3} q^2\big)/V^{2/3}-3}{9 \pi  TV}.
\end{equation}
%On the \((T,V)\)-plane, the Ruppeiner line element is:
%\begin{equation}
%ds_R^2 = \left( \frac{-4 (6 \pi )^{2/3} q^2-3 (6 \pi )^{2/3} T V+3 \sqrt[3]{6} V^{2/3}}{54 \sqrt[3]{\pi } T V^{7/3}}  \right)\,dV^2+ \frac{2\sqrt[3]{\frac{\pi }{6}}}{T \sqrt[3]{V}}\, dT \, dV
%\end{equation}
%The corresponding Ricci scalar is given by,
The curvature scalar is plotted in figure-(\ref{R_V_RNAdS}) as a function of \(V\) for fixed \(T\). We may now probe the nature of interactions between black hole microstructures. It is straightforward to see that the curvature is positive and hence repulsive interactions dominate if for a fixed electric charge, the horizon radius satisfies the following bound upper bound $r_+ < 1.414213|q|$.
\noindent
Beyond this upper bound, the Ruppeiner curvature is negative. Therefore, for larger RN-AdS black holes with the same electric charge, i.e. with the corresponding volume beyond this upper bound there is a dominance of attractive interactions among the black hole microstructures.\\
\begin{figure}[h]
%	 \begin{wrapfigure}{l}{0.3\textwidth}
	\begin{center}
		\centering
		\includegraphics[width=4.6in]{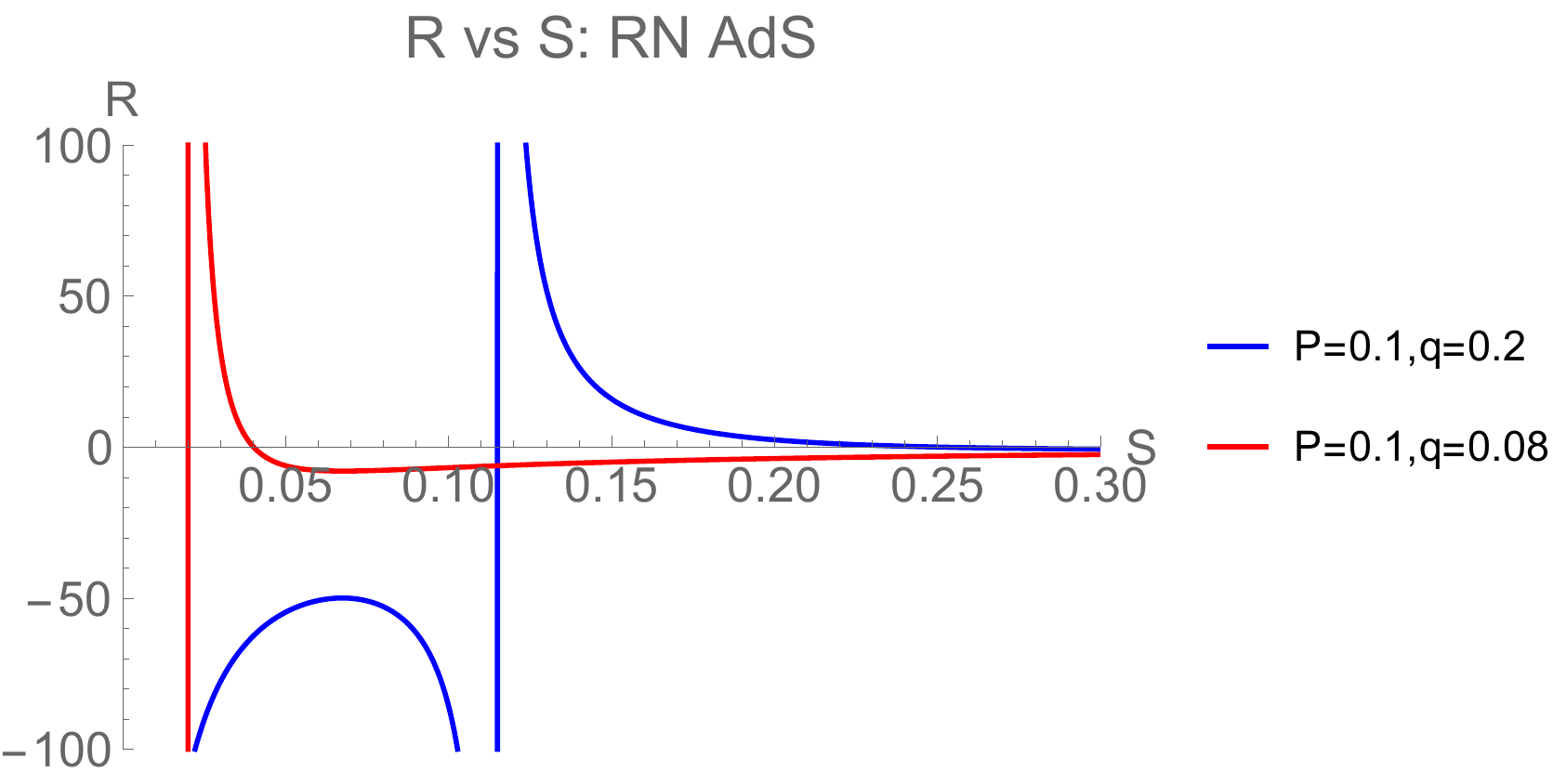}  		
		\caption{Curvature of Ruppeiner metric on \((S,P)\)-plane for RN-AdS black holes as a function of \(S\) for fixed \(P\).}   \label{R_S_RNAdS}	
        \includegraphics[width=4.6in]{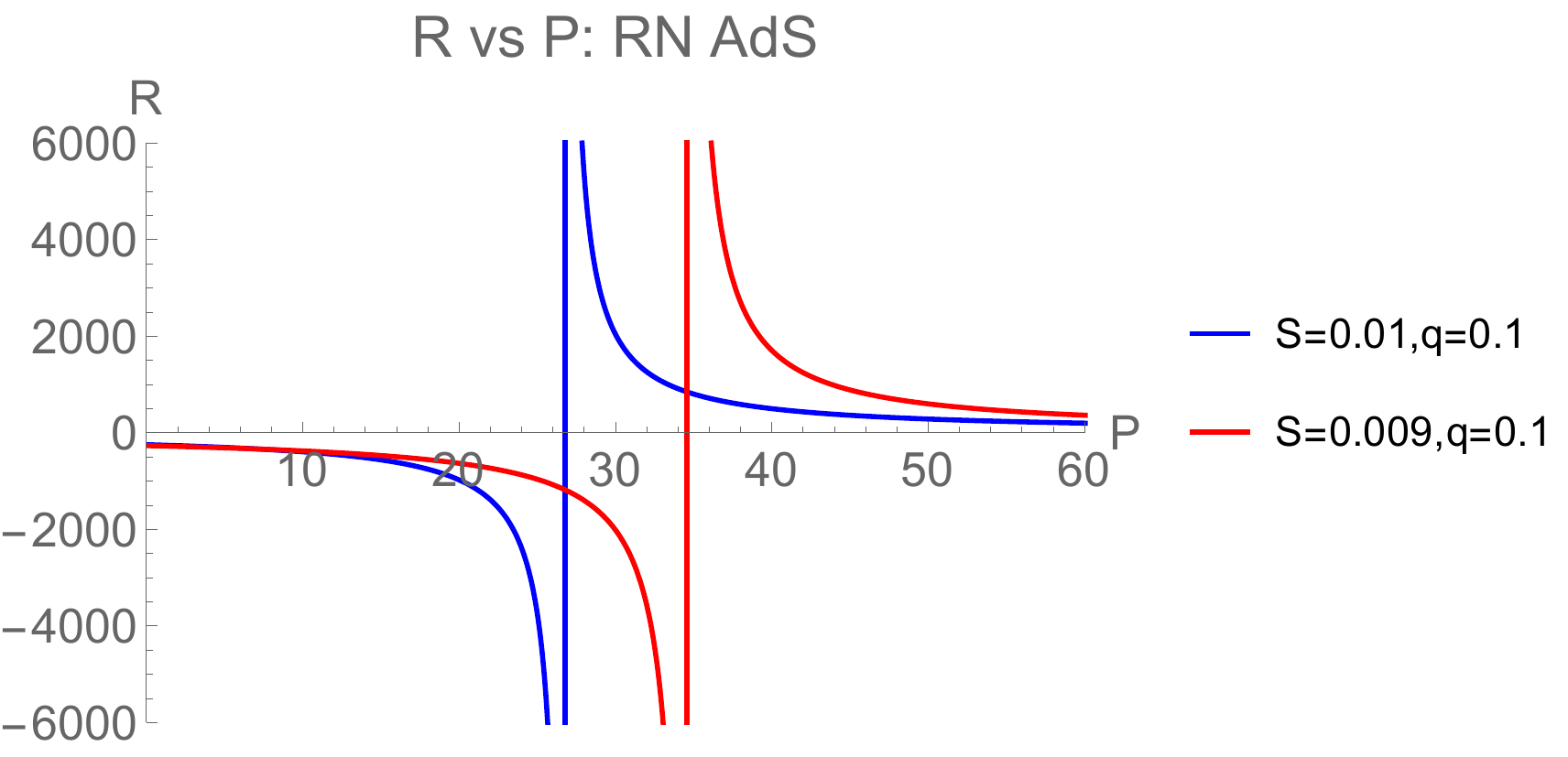}  		
		\caption{Curvature of Ruppeiner metric on \((S,P)\)-plane for RN-AdS black holes as a function of \(P\) for fixed \(S\).}   \label{R_P_RNAdS}	
	\end{center}
%	\end{wrapfigure}
\end{figure}
\begin{figure}[h]
%	 \begin{wrapfigure}{l}{0.3\textwidth}
	\begin{center}
		\centering
		\includegraphics[width=4.6in]{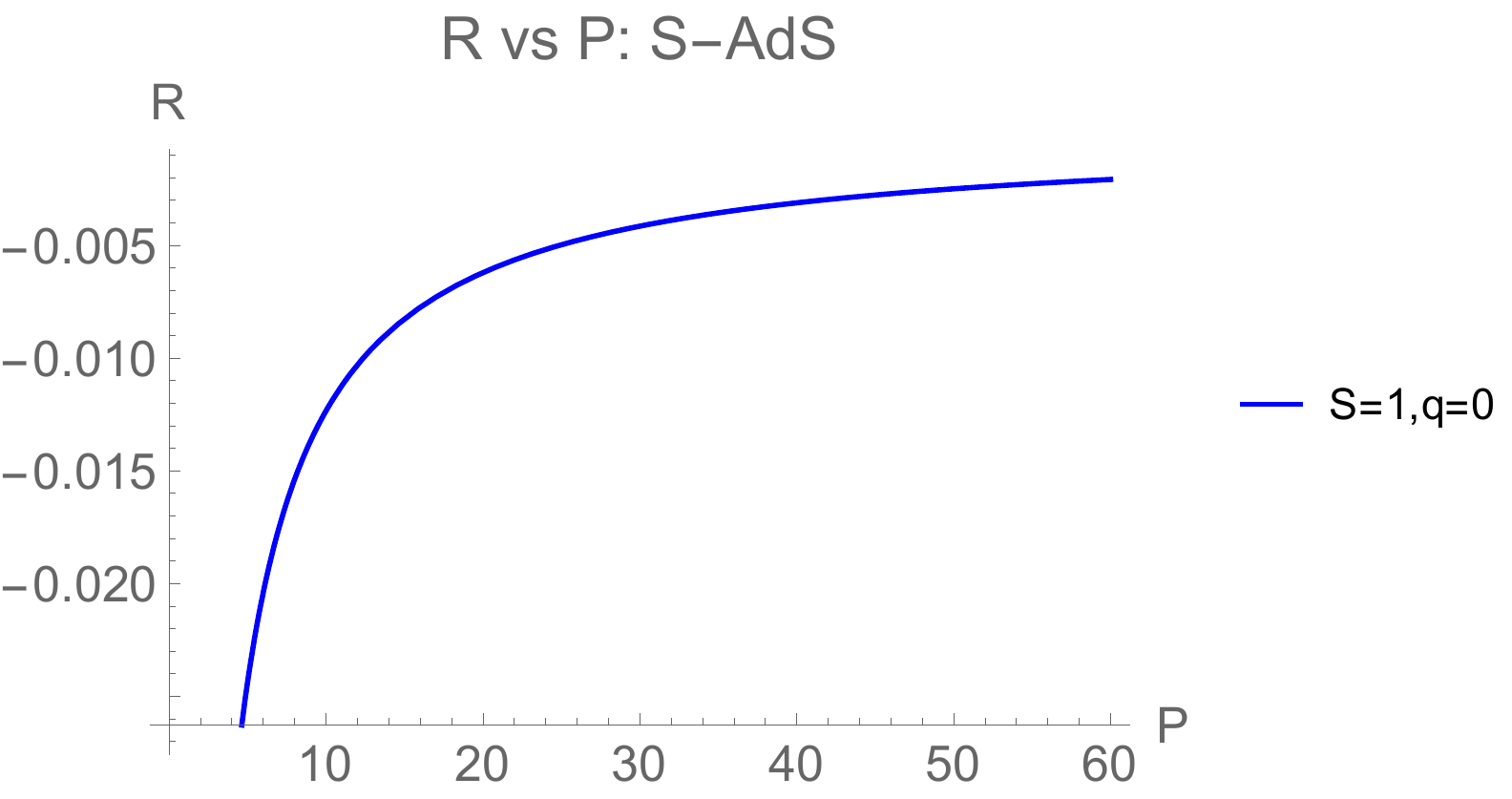}  		
		\caption{Curvature of Ruppeiner metric on \((S,P)\)-plane for Schwarzschild-AdS black holes as a function of \(P\) for fixed \(S\).}   \label{R_P1_RNAdS}	
	\end{center}
%	\end{wrapfigure}
\end{figure}
\noindent
Now considering the \((S,P)\)-plane, the corresponding Ricci scalar is computed from eqn (\ref{RuppeinerSP}) to be,
%Now consider the \((S,P)\)-plane. In this case, the metric takes the form:
%\begin{equation}
%ds_R^2 = \left( \frac{8 P S^2+3 \pi  q^2-S}{2 S \left(8 P S^2-\pi  q^2+S\right)}  \right)\,dS^2+ \frac{16 S^2}{8 P S^2-\pi  q^2+S}\, dS \, dP
%\end{equation}
%with the corresponding Ricci scalar,
\begin{equation}\label{RN_R_SP}
R_{SP} =-\frac{S-2 \pi  q^2}{S \left(8 P S^2-\pi  q^2+S\right)}.
\end{equation}
It can be checked explicitly that the scalar curvatures on both the representations are equivalent and hence we relabel them as \(R\). The curvature scalar has been plotted in figure-(\ref{R_S_RNAdS}) as a function of \(S\) for fixed \(P\) and in figure-(\ref{R_P_RNAdS}) as a function of \(P\) for fixed \(S\). Yet again, it is easy to check from eqn (\ref{RN_R_SP}) that the curvature is positive for black holes with \(r_+\) satisfying the bound given by $r_+ < 1.414213|q|$. We note that the metric is flat if,
\begin{equation}\label{RNAdSNoninteracting}
  r_+ = 1.414213|q|
\end{equation}
This is the case where the attractive and repulsive interactions balance each other exactly so that there are no net interactions between the black hole microstructures. This can also be seen as a stationary point of the pressure of the RN-AdS black hole [eqn (\ref{RNeqnofstatev})] in the absence of thermal excitations,
\begin{equation}
  \bigg(\frac{\partial P}{\partial v}\bigg)_{T=0} = 0
\end{equation} giving \(r_+ = 0.5v = 1.414213|q|\). For \(T=0\), no kinetic pressure remains, so that the entire pressure is due to interactions and hence, this point should be interpreted as the point where the black hole transitions from a repulsion dominated region to an attraction dominated region and vice versa. This is however, not a true phase transition of black hole since the kinetic effects due to a thermal background are not taken into account.\\

\noindent
It should also be remarked that if we take the electric charge to be zero then all the results shall correspond to those for Schwarzschild black holes in AdS\(_4\) \cite{SAdS} with the scalar curvatures of the Ruppeiner metrics on both \((T,V)\)- and \((S,P)\)-planes reducing respectively to,
\begin{equation}\label{S-AdS}
  R = \frac{-1}{3 \pi  TV} = \frac{-1}{S(8 P S+1)}.
\end{equation}
The curvature is negative indicating attractive interactions among black hole microstructures in the Schwarzschild-AdS black hole \cite{SAdS}. This can also be seen from eqn (\ref{RNeqnofstatev}) where in this case with \(q=0\) the interactions are purely attractive. The Ricci scalar is plotted as a function of \(P\) holding \(S\) fixed for \(q=0\) in figure-(\ref{R_P1_RNAdS}), which corresponds to the Schwarzschild-AdS case. Although, there are other divergences and points where curvature takes negative values in figures-(\ref{R_S_RNAdS}) and (\ref{R_P_RNAdS}), these can be ignored from thermodynamic considerations by imposing the positivity of temperature. The curves for curvature, on the far right in these figures, which start out as positive, cross zero and become negative are thus the only ones considered~\cite{Mann2019,Wei:2019ctz}.

%%%%%%%%%%%%%%%%%%%%%%%%%%%%%%%%%%%%
\section{Charged Gauss-Bonnet black holes in AdS\(_5\)} \label{GBAdSBH}
%%%%%%%%%%%%%%%%%%%%%%%%%%%%%%%%%%%%%%
The Einstein-Maxwell action in a \(5\)-dimensional AdS background with a Gauss-Bonnet term can be written as~\cite{Cai:2001dz,GB},
\begin{equation}\label{ActionGB}
  \mathcal{S} = \frac{1}{16\pi}\int d^5x \sqrt{-g}[R - 2\Lambda + \alpha_{GB}(R_{\mu \nu \gamma \delta}R^{\mu \nu \gamma \delta} - 4 R_{\mu \nu}R^{\mu \nu} + R^2) - 4\pi F^{\mu \nu}F_{\mu \nu}],
\end{equation} with \(\alpha_{GB}\) being the Gauss-Bonnet coupling constant, \(F_{\mu \nu}\) the electromagnetic field tensor and the cosmological constant \(\Lambda\) being given as,
\begin{equation}
  \Lambda = -\frac{(d-2)(d-1)}{2l^2},
\end{equation}
where the constant \(l^2\) measures the curvature of the AdS spacetime. The mass \(M\) of the black hole is obtained to be,
\begin{equation} \label{massGB}
M=\frac{(d-2)\omega_{d-2}}{16\pi}\left(\alpha r_+^{d-5}+r_+^{d-3}+\frac{q^2}{r_+^{d-3}}+16\pi P \frac{r_+^{d-1}}{(d-2)(d-1)}\right) ,
\end{equation}
and the entropy as a function of the horizon radius (for $d=5$) is expressed as,
\begin{equation}
S = \frac{\pi ^2 r_+^3}{2}+3 \pi ^2 \alpha  r_+\, ,
\end{equation}
which on solving for \(r_+\) one gets,
\begin{equation}\label{r_+(S)}
r_+=\frac{\left(\sqrt{8 \pi ^4 \alpha ^3+S^2}+S\right)^{2/3}-2 \pi ^{4/3} \alpha }{\pi ^{2/3} \sqrt[3]{\sqrt{8 \pi ^4 \alpha
   ^3+S^2}+S}}.
\end{equation}
Therefore from eqns (\ref{massGB}) and (\ref{r_+(S)}), the mass of a charged Gauss-Bonnet-AdS (GB-AdS) black hole equated to the enthalpy in $d=5$, is expressed as,
\begin{eqnarray}\label{HGB}
&& H(S,P) = \frac{3}{8} \pi \left( h_1+h_2+h_3+\alpha \right), \nonumber \\
&& h_1 = \frac{\pi ^{4/3} q^2 \left(\sqrt{8 \pi ^4 \alpha ^3+S^2}+S\right)^{2/3}}{\left(\left(\sqrt{8 \pi ^4 \alpha ^3+S^2}+S\right)^{2/3}-2 \pi ^{4/3}
   \alpha \right)^2} \, ,\quad
 h_2=  \frac{4 P \left(\left(\sqrt{8 \pi ^4 \alpha ^3+S^2}+S\right)^{2/3}-2 \pi ^{4/3} \alpha \right)^4}{3 \pi ^{5/3} \left(\sqrt{8 \pi ^4 \alpha
   ^3+S^2} +S\right)^{4/3}}, \nonumber \\
&& h_3= \frac{\left(\left(\sqrt{8 \pi ^4 \alpha ^3+S^2}+S\right)^{2/3}-2 \pi ^{4/3} \alpha \right)^2}{\pi ^{4/3} \left(\sqrt{8 \pi ^4 \alpha
   ^3+S^2}+S\right)^{2/3}}.
\end{eqnarray}
\noindent
The equation of state of the Gauss-Bonnet-AdS black hole in general dimensions with constant electric charge can be obtained to be,
\begin{equation} \label{eosd}
P=-\frac{(d-2) r_+^{-2 d-4}}{16 \pi } \left(-d q^2 r_+^8+\alpha  d r_+^{2 d}-5 \alpha  r_+^{2 d}-8 \pi  \alpha  T r_+^{2 d+1}-4 \pi  T r_+^{2 d+3}+d
   r_+^{2 d+2}-3 r_+^{2 d+2}+3 q^2 r_+^8\right),
\end{equation}
where \(\alpha = (d-3)(d-4)\alpha_{GB}=2\alpha_{GB}\) and the topology of the black hole horizon is taken to be spherical. In $d=5$, the equation of state is,
\begin{equation}\label{GBeqnr}
P = \frac{3 \left(q^2+2 \pi  r_+^5 T-r_+^4+4 \pi  \alpha  r_+^3 T\right)}{8 \pi  r_+^6} \, .
\end{equation}
Now using the relation $r_+=\frac{\sqrt[4]{2} \sqrt[4]{V}}{\sqrt{\pi }}$, one can write,
\begin{equation}\label{GBeqnV}
P=\frac{3 \left(\pi ^2 q^2+4\ 2^{3/4} \pi ^{3/2} \alpha  T V^{3/4}+4 \sqrt[4]{2} \sqrt{\pi } T V^{5/4}-2 V\right)}{16 \sqrt{2}
   V^{3/2}} \, .
\end{equation}
Moreover, using \(v = 4r_+/3\) for specific volume of the black hole in \(d=5\), the equation of state can be expressed in terms of the specific volume from eqn (\ref{GBeqnr}) as,
 \begin{equation}\label{GBeqnofstate}
  P = \frac{T}{v}\bigg( 1 + \frac{32\alpha}{9 v^2}\bigg) - \frac{2}{3 \pi v^2} + \frac{512q^2}{243 \pi v^6} \, ,
\end{equation}
where one can see that interaction terms modify the pressure similar to the case for RN-AdS black holes. \\

\begin{figure}[h]
%	 \begin{wrapfigure}{l}{0.3\textwidth}
	\begin{center}
		\centering
		\includegraphics[width=4.6in]{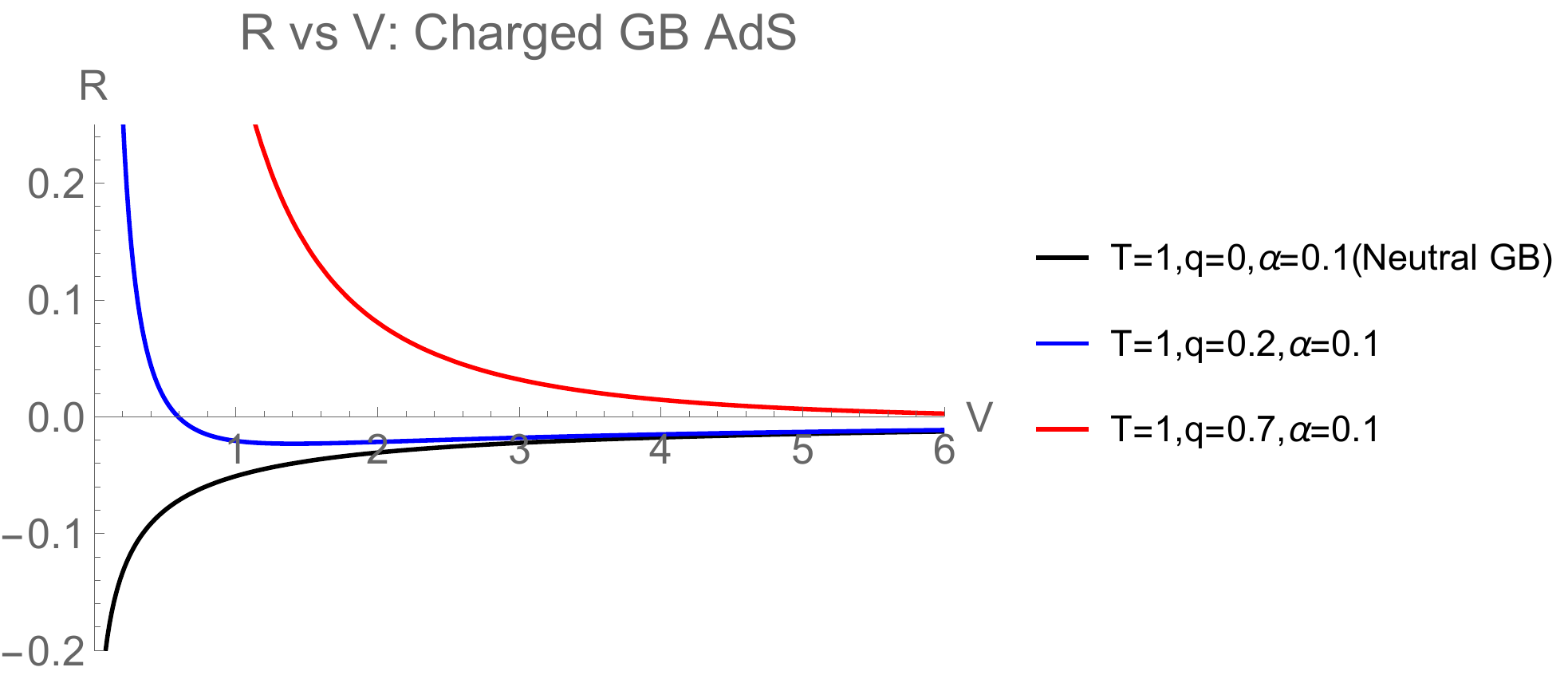}  		
		\caption{Curvature of Ruppeiner metric on \((T,V)\)-plane for Gauss-Bonnet-AdS black holes as a function of \(V\) for fixed \(T\).}   \label{R_V_GBAdS}		
	\end{center}
%	\end{wrapfigure}
\end{figure}

\noindent
Now, for the \(d=5\) case, the Ruppeiner metric on the $(T,V)$-plane can be calculated easily from eqns (\ref{GBeqnV}) and (\ref{RuppeinerTV}). The corresponding Ricci scalar is obtained to be,
\begin{equation}\label{curvature_GB_TV}
R_{TV} =-\frac{2 V-3 \pi ^2 q^2}{6 \pi  T V \left(\sqrt{2} \pi  \alpha +\sqrt{V}\right)^2}.
\end{equation}
Therefore, for charged Gauss-Bonnet-AdS black holes with volumes below the given bound, the interactions between the microstructures are dominantly repulsive. A similar feature was observed for the case of RN black holes. On the other hand, for black holes with horizon radii, \(r_+ > 1.316074 |q|^{0.5}\), the Ruppeiner curvature changes sign to negative indicating the dominance of attractive interactions. These features can be seen in figure-(\ref{R_V_GBAdS}) where the curvature has been plotted as a function of volume for fixed temperature. The curvature vanishes at,
\begin{equation} \label{boundGB}
               r_+ = 1.316074 |q|^{0.5},
             \end{equation}
for which the attractive and repulsive interactions balance each other.
This point can also be determined as the physical solution of the equation,
\begin{equation}
  \bigg(\frac{\partial P}{\partial v}\bigg)_{T=0} = 0,
\end{equation}
where \(P\) is given by eqn (\ref{GBeqnofstate}) showing that there is a transition from the repulsion dominated region to the attraction dominated region at this point.  This too is not a true phase transition since thermal effects have not been considered. The point \(T=0\) for the temperature is chosen to extinguish all thermal effects so that only the effects due to the intrinsic properties of the microstructures remain. In this context, the interaction term \(32\alpha T/9 v^3\) can be viewed only as a pseudo-interaction that activates when there is a thermal background and does not control in any manner, the sign of the Ruppeiner curvature as can be seen from eqn (\ref{curvature_GB_TV}).
It is then easy to check from eqn (\ref{curvature_GB_TV}) that the Ruppeiner curvature is positive for black holes with their size bounded by the inequality, $r_+ < 1.316074 |q|^{0.5}$.
For the charge neutral case, i.e. for \(q=0\), it is seen that the curvature is always negative indicating that attractive interactions dominate, matching the results obtained for the case of neutral Gauss-Bonnet-AdS black holes in~\cite{Wei:2019ctz}.\\

\begin{figure}[h]
%	 \begin{wrapfigure}{l}{0.3\textwidth}
	\begin{center}
		\centering
		\includegraphics[width=4.6in]{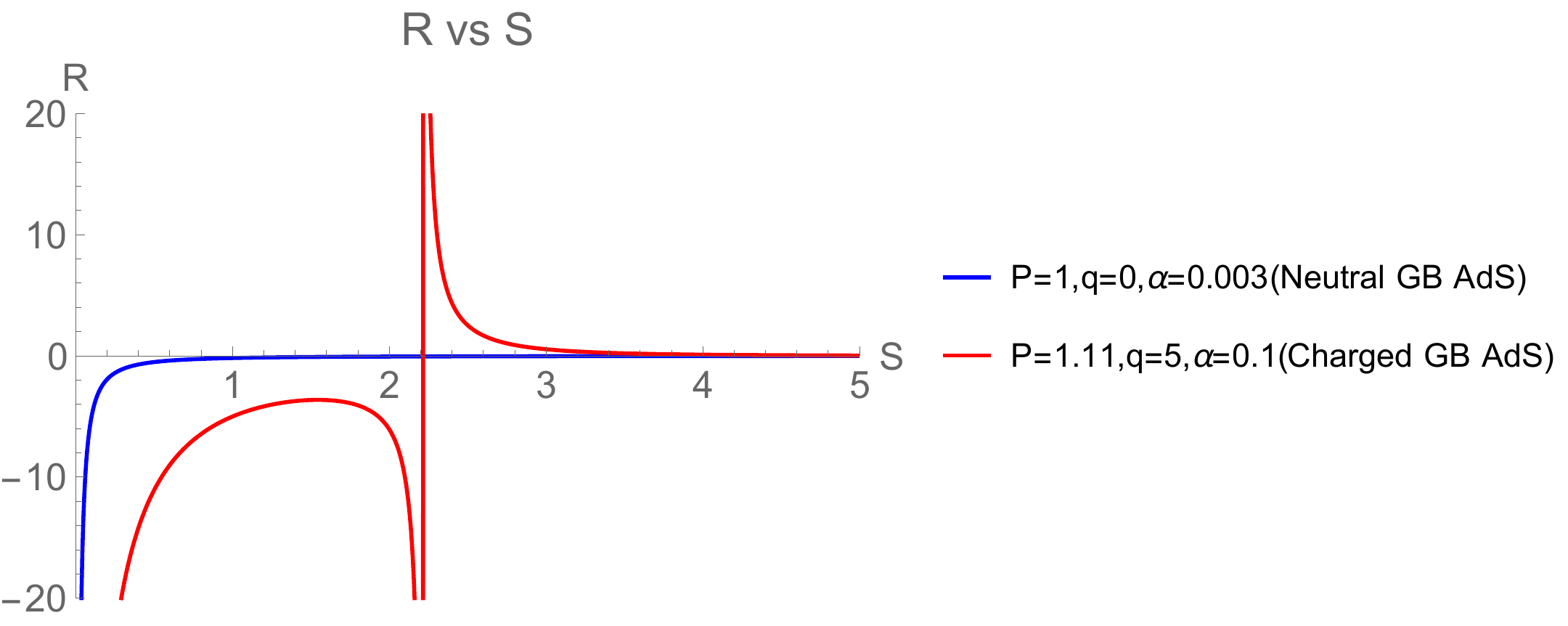}  		
		\caption{Curvature of Ruppeiner metric on \((S,P)\)-plane for Gauss-Bonnet-AdS black holes as a function of \(S\) for fixed \(P\).}   \label{R_S_GBAdS}		
	\end{center}
%	\end{wrapfigure}
\end{figure}

\noindent
The Ruppeiner metric in $(S,P)$-plane can be written down analytically. We record the Ricci scalar to be,
\begin{equation}
R_{SP} = -\frac{A}{B},
\end{equation}
where,
\begin{eqnarray}
&& A = 8 \pi  \left(\sqrt{8 \pi ^4 \alpha ^3+S^2}+S\right)^{8/3} \left(S-\frac{2 \pi ^{4/3} \alpha  \left(\left(\sqrt{8 \pi ^4
   \alpha ^3+S^2}+S\right)^{2/3}-2 \pi ^{4/3} \alpha \right)}{\sqrt[3]{\sqrt{8 \pi ^4 \alpha ^3+S^2}+S}}\right) \nonumber \\
   && \qquad \qquad \times
   \left(\frac{\left(\left(\sqrt{8 \pi ^4 \alpha ^3+S^2}+S\right)^{2/3}-2 \pi ^{4/3} \alpha \right)^4}{\pi ^{2/3}
   \left(\sqrt{8 \pi ^4 \alpha ^3+S^2}+S\right)^{4/3}}-3 \pi ^2 q^2\right), \nonumber
 \end{eqnarray}
 and
 \begin{eqnarray}
&& B=8 P \left(\left(\sqrt{8 \pi ^4 \alpha ^3+S^2}+S\right)^{2/3}-2 \pi ^{4/3} \alpha \right)^6-3 \pi ^3 q^2 \left(\sqrt{8 \pi ^4
   \alpha ^3+S^2}+S\right)^2 \nonumber \\
   &&+3 \sqrt[3]{\pi } \left(\sqrt{8 \pi ^4 \alpha ^3+S^2}+S\right)^{2/3} \left(\left(\sqrt{8 \pi ^4
   \alpha ^3+S^2}+S\right)^{2/3}-2 \pi ^{4/3} \alpha \right)^4 \nonumber \\
&&\left(\left(\sqrt{8 \pi ^4 \alpha ^3+S^2}+S\right)^{2/3}-2 \pi ^{4/3} \alpha \right)^2
\left(2 \pi ^{4/3} \alpha +\sqrt{\frac{\left(\left(\sqrt{8 \pi ^4 \alpha ^3+S^2}+S\right)^{2/3}-2 \pi ^{4/3} \alpha
   \right)^4}{\left(\sqrt{8 \pi ^4 \alpha ^3+S^2}+S\right)^{4/3}}}\right)^2. \nonumber
\end{eqnarray}
The Ricci scalar has been plotted as a function of \(S\) for fixed \(P\) in figure-(\ref{R_S_GBAdS}). It is seen that for the charge neutral case, the curvature is always negative whereas, the curvature can be both positive or negative for the electrically charged case depending upon the value of \(r_+\). Again, only the curve for curvature, on the far right  in figure-(\ref{R_S_GBAdS}), which start out as positive, crosses zero and become negative is physical from thermodynamics point of view, as temperature is positive only for this curve.
A physical understanding of the attraction-repulsion regimes can be achieved by the following simple model. In this picture where black holes are associated with microstructures taking up microscopic degrees of freedom, one may associate two distinct classes of microstructures \cite{AR} to charged black holes. The first type is the charge neutral type while the second type corresponds to charged microstructures. The microstructures of the first class, i.e. those without electric charge are characterized by attractive interactions between them as is the case of gas molecules in the van der Waals fluid while the charged microstructures falling in the second class interact with each other in a repulsive manner. Neutral black holes such as the Schwarzschild-AdS and the neutral Gauss-Bonnet-AdS black holes can be understood to be comprised of only neutral microstructures and hence only attractive interactions prevail as is seen from the negative sign of the Ruppeiner curvature. On the other hand, the RN-AdS and the charged GB-AdS black holes are comprised of microstructures from both the classes like a binary mixture of two fluids. This results in a competition among the repulsive and attractive interactions. The electric charge of a black hole is proportional to the number of charged and hence repulsive black hole microstructures. Therefore, for a fixed value of the electric charge and hence, a fixed number of repulsive microstructures the effective repulsion or attraction is controlled by the number of neutral (attractive) microstructures. For example in the Gauss-Bonnet in AdS case, if we take a particular value of charge \(q_0\) and fix it, the number of charged microstructures is held fixed and the remaining microscopic degrees of freedom are therefore associated with the neutral microstructures. Hence, for Gauss-Bonnet-AdS black holes that have entropy (and hence horizon radius) below a certain value, the neutral microstructures are outnumbered by their charged counterparts leading to an overall dominance of repulsive interactions in the black hole. This upper bound in the horizon radius is expressed in this case as \(r_+ < 1.316074 |q_0|^{0.5}\). For black holes with larger values of entropy corresponding to larger horizon radii, if the electric charge is still held fixed at \(q_0\), the neutral microstructures outnumber the charged ones leading to the domination of attractive interactions.  If both the effects effectively cancel, the Ruppeiner metric is rendered flat. This happens for \(r_+ = 1.316074 |q_0|^{0.5}\) for the charged Gauss-Bonnet-AdS black holes in $d=5$. This same feature is seen for the RN-AdS case  in $d=4$ with the corresponding point at which the interactions balance being given as\footnote{In $d=5$: The bound on the horizon radius for GB-AdS and RN-AdS black hole cases, are found to be exactly identical.} \(r_+ = 1.414213|q_0|\).  The domination of attractive attractions for neutral Gauss-Bonnet-AdS black holes has recently been noted in~\cite{Wei:2019ctz}, but with a different metric than the one presented here\footnote{The metrics used in~\cite{Mann2019,Wei:2019ctz,Wei:2019yvs} to discuss the thermodynamic geometry of RN-AdS and neutral GB-AdS black holes assume that $C_V \neq 0$.  Following~\cite{Mann2019,Wei:2019ctz,Wei:2019yvs}, if we assume $C_V \neq 0$, then our metric in $(T,V)$-plane turns out to be (which is still different from the one used in~\cite{Mann2019}),
\begin{equation}\label{RuppeinerTVCv}
  ds_R^2 =\frac{C_V}{T^2} \, dT^2 + \frac{1}{T}\bigg(\frac{\partial P}{\partial V}\bigg)_TdV^2 + \frac{2}{T}\bigg(\frac{\partial P}{\partial T}\bigg)_VdTdV \, ,
\end{equation}
with the corresponding curvature for charged GB-AdS black holes calculated to be,
\begin{equation} \label{curvGB}
R_{Tv}=\frac{512 \left(256 q^2-27 v^4\right) \left(32768 q^2+27 v^3 \left(2048 \pi  \alpha  T+81 \pi ^3 T v^5+576 \pi ^3 \alpha  T
   v^3+192 \pi  T v^2+128 v \left(8 \pi ^3 \alpha ^2 T-1\right)\right)\right)}{3 \left(65536 q^2+27 v^3 \left(2048 \pi
   \alpha  T+81 \pi ^3 T v^5+576 \pi ^3 \alpha  T v^3+192 \pi  T v^2+256 v \left(4 \pi ^3 \alpha ^2
   T-1\right)\right)\right)^2} \, .
\end{equation}
Here, the curvature scalar in eqn (\ref{curvGB}) is expressed in terms of the specific volume \(v\) rather than the thermodynamic volume \(V\). There are of course new divergence points of the curvature in eqn (\ref{curvGB}), but they can be removed from thermodynamic considerations, using the coexistence curves; with discussion being similar to the cases studied in~\cite{Mann2019,Wei:2019ctz,Wei:2019yvs}, and we do not pursue it here. More importantly, the qualitative features are similar to the RN AdS case noted in~\cite{Wei:2019ctz} (see figure-(11) in page 25); at smaller values of $v$, the curvature starts out as positive (signifying repulsive interactions) and then goes to negative values (attractive interactions). Thus, although the curvature in eqn (\ref{curvGB}) is different from the one in eqn (\ref{curvature_GB_TV}), it still has the same bound (obtained from $\left(256 q^2-27 v^4\right)=0$ and using the fact that $v=4r_+/3$) as noted in eqn (\ref{boundGB}), and associated features discussed thereafter. Thus, as far as probing the interaction among microstructures of black holes are concerned, the metric proposed in eqns (\ref{RuppeinerSP}) and (\ref{RuppeinerTV}) might be more useful, as the Ruppeiner curvature is more simpler in this parameterization.}. Furthermore, the interplay of repulsive and attractive forces, and connections to fermion and boson gases have been noted in~\cite{Wei:2012ui}, though the computations were in non-extended phase space, using a different thermodynamic potential for Gauss-Bonnet black holes.\\

\noindent
We now comment on our results in other dimensions. As the computations are similar in nature, we directly note the bound found on the horizon radius.
In $d=6$, we find the following bound where the attractive and repulsive interactions balance each other,
\begin{figure}[h]
%	 \begin{wrapfigure}{l}{0.3\textwidth}
	\begin{center}
		\centering
		\includegraphics[width=3.0in]{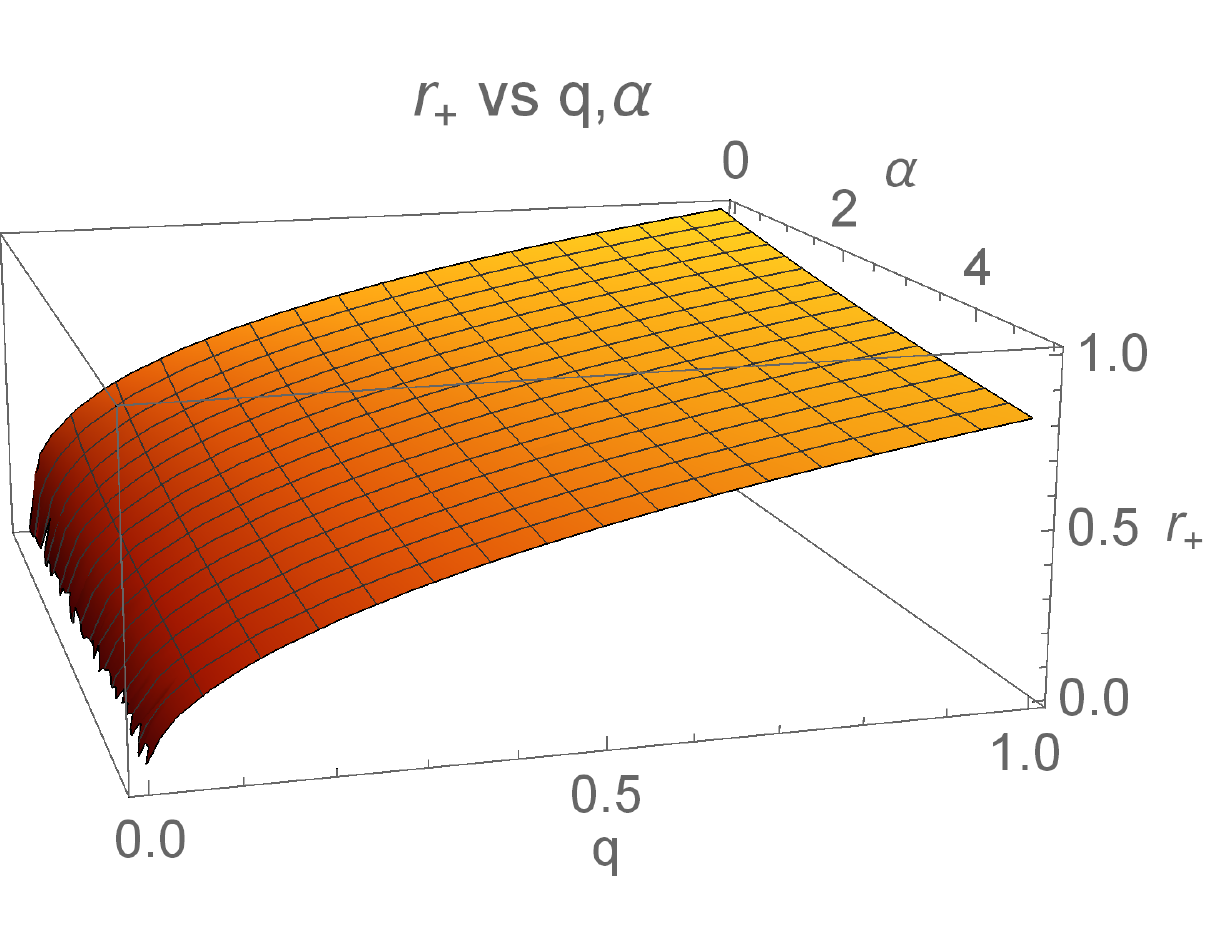}  		
		\caption{Horizon radius $r_+$ plotted against charge $q$ and GB coupling $\alpha$ in $d=6$: for the points where the Ruppeiner curvature is zero - indicating the balance of attractive and repulsive interactions.}   \label{r_q_alpha_6D}	
	\end{center}
%	\end{wrapfigure}
\end{figure}
\begin{equation}
r_+ =\sqrt{-\frac{\alpha }{9}+\frac{\sqrt[3]{-2 \alpha ^3+729 q^2+27 \sqrt{729 q^4-4 \alpha ^3 q^2}}}{9
   \sqrt[3]{2}}+\frac{\sqrt[3]{2} \alpha ^2}{9 \sqrt[3]{-2 \alpha ^3+729 q^2+27 \sqrt{729 q^4-4 \alpha ^3 q^2}}}} \, .
\end{equation}
Unlike in $d=5$, this bound now depends on the GB coupling $\alpha$, apart from its dependence on charge $q$. These dependencies are plotted in figure-(\ref{r_q_alpha_6D}), which reveals that bound on horizon radius $r_+$ increases with $q$ and decreases with $\alpha$. We have checked that this feature continues to hold in higher dimensions too.

%%%%%%%%%%%%%%%%%%%
\section{Remarks}
%%%%%%%%%%%%%%%%%%%%%

%In this paper we studied Ruppeiner geometry for \(d=5\) charged Gauss-Bonnet black holes in an AdS background. The identification of thermodynamic pressure with the cosmological constant leads to the inclusion of \(P\) and \(V\) terms in black hole thermodynamics. However, it turns out that for static black holes the thermodynamic volume is not independent of the black hole entropy, both being functions of the horizon radius. For example, in case of RN black holes in AdS\(_4\), the volume and entropy are respectively given as:
%$$ V = \frac{4}{3}\pi r_+^3; \hspace{3mm} S = \pi r_+^2 $$
%clearly showing that they are not independent. Therefore, as suggested in \cite{SAdS}, Ruppeiner geometry on the \((S,V)\)-plane or an internal energy \(U = U(T,V)\) representation is not an appropriate setting. Instead, we study Ruppeiner geometry on the \((T,V)\) and \((S,P)\) phase planes.

%\smallskip \noindent

\begin{figure}[h]
%	 \begin{wrapfigure}{l}{0.3\textwidth}
	\begin{center}
		\centering
		\includegraphics[width=4.6in]{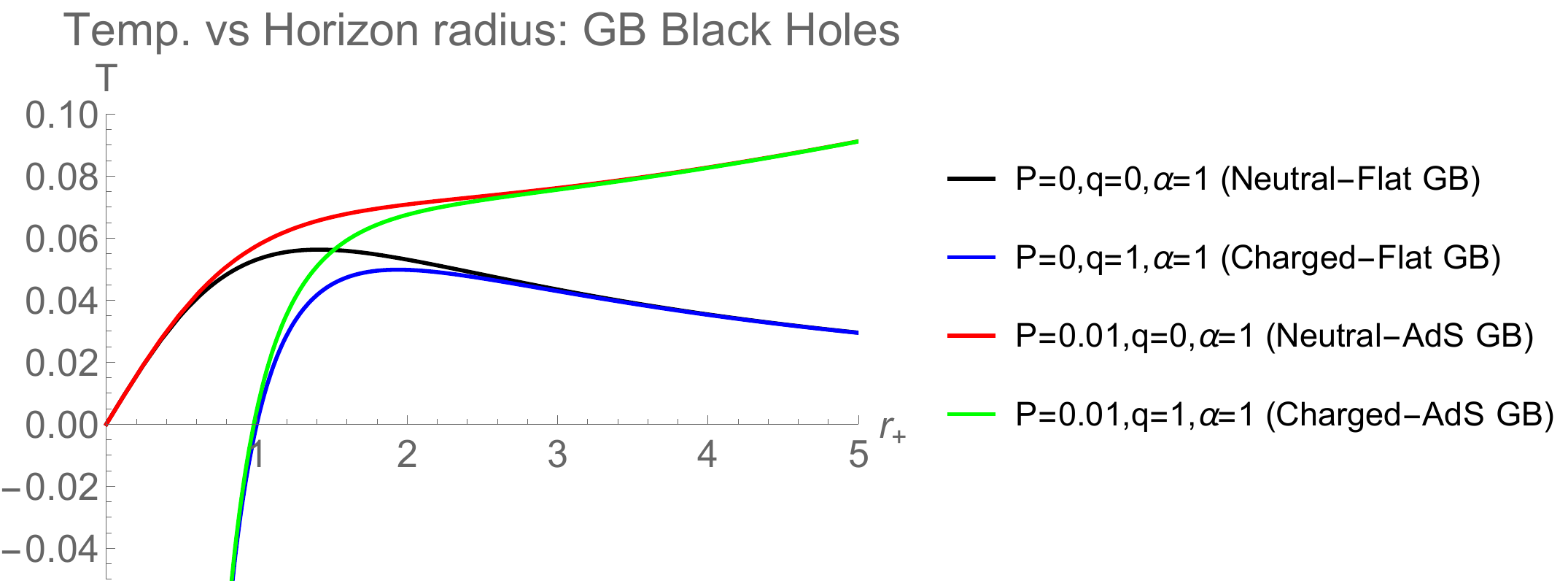}  		
		\caption{Temperature vs horizon radius for Gauss-Bonnet black holes.}   \label{T_r_GBAdS}		
	\end{center}
%	\end{wrapfigure}
\end{figure}

Because black holes can be associated with a temperature, it is therefore natural to associate a microscopic description to black holes in terms of the so called black hole microstructures. In this description, Ruppeiner geometry turns out to be an extremely useful method to probe the nature of interactions between these black hole microstructures. We found that for RN black holes in AdS\(_4\) with their size limited by the upper bound, \(r_+ < 1.414213|q|\) where \(q\) is the electric charge, the repulsive interactions dominate whereas for the black holes larger than this bound, the attractive interactions dominate. If however, the horizon radius satisfies \(r_+ = 1.414213|q|\), the attractive and repulsive interactions balance exactly. Since Schwarzschild-AdS black holes are the electrically neutral counterparts, it is immediately clear that for the Schwarzschild-AdS case only attractive interactions are dominant. Similar results were observed with charged Gauss-Bonnet black holes in AdS\(_5\) with the upper bound in the horizon radius below which repulsive interactions dominate is \(r_+ < 1.316074 |q|^{0.5}\). In five dimensions, the bounds on $r_+$ for RN-AdS and GB-AdS black holes were found to be identical. In higher dimensions however, we found that the bound on $r_+$ depends on $\alpha$ too. In particular bound on $r_+$ increases with charge $q$ and decreases with $\alpha$.  This bound on $r_+$ is physically explained by considering two distinct classes of black hole microstructures sharing the microscopic degrees of freedom of the black hole entropy. Charged black holes are taken to be constituted by microstructures of both the classes in different proportions just like a two fluid mixture leading to the possibility of both attractive or repulsive interactions depending on which class of microstructures dominate. Neutral black holes however are associated with only neutral microstructures leading to attractive interactions in their microstructures.
\smallskip
\noindent
Since charged black holes are regarded as a binary mixture of two fluids respectively comprising of neutral and charged particles, it follows that for a given fixed value of electric charge and hence, for a fixed number of charged microstructures, the larger the size of the black hole the more dominating are thermodynamic effects due to the uncharged counterparts. This is easily seen from the variation of the black hole's temperature with the horizon radius which for charged Gauss-Bonnet black holes in AdS\(_5\) is expressed as,
\begin{equation}\label{T GB}
T = \frac{\frac{16}{3} \pi  P r_+^4-\frac{2 q^2}{r_+^2}+2 r_+^2}{4 \pi  r_+^3+8 \pi  \alpha  r_+}.
\end{equation}
The variation of the temperature with horizon radius is shown in figure-(\ref{T_r_GBAdS}). It is immediately clear that even though for smaller sizes, the temperatures for both charged and neutral Gauss-Bonnet black holes in AdS background differ significantly but are close to those for the respective flat space black holes; for the larger black holes, the temperatures for charged and neutral Gauss-Bonnet black holes are similar in both AdS or flat space backgrounds. One can therefore say that the thermodynamic effects due to the presence of charged microstructures are essentially overwhelmed by the neutral microstructures for large sized black holes. This simply amounts to dropping the term with electric charge in the numerator of eqn (\ref{T GB}). This is also observed from the expressions of the Ruppeiner curvatures. Since, the black hole entropy \(S\) is an increasing function of the horizon radius, it follows that in the large \(r_+\) limit the entropy is so large enough that the terms containing electric charge in the Ruppeiner curvature have insignificant contributions.

\smallskip \noindent
As has been remarked in \cite{AR}, electric charge has a fundamental role in phase transitions of black holes and Ruppeiner geometry proves to be instrumental in revealing such critical aspects. Furthering the idea of black hole microstructures shall allow the generalization of several results of kinetic molecular theory to black hole physics. It would be very interesting to see the attraction-repulsion interplay for rotating black holes where it might be suggestive to assign a non-zero angular momentum to the black hole microstructures.

\appendix
\section*{Appendices}
\addcontentsline{toc}{section}{Appendices}
\renewcommand{\thesubsection}{\Alph{subsection}}

\subsection{Contact Hamiltonian dynamics}
\renewcommand{\theequation}{\thesubsection.\arabic{equation}}
It is now understood that the thermodynamic phase space is a contact manifold \((\mathcal{M},\eta)\). We may therefore define contact Hamiltonian dynamics on the thermodynamic phase space. For an arbitrary differentiable function, \(h:\mathcal{M} \rightarrow \R\) we may associate a vector field \(X_h\) defined from \(h\) by the following conditions,
\begin{equation}\label{contactvectorfield}
 i_{X_h}\eta = -h, \hspace{3mm} i_{X_h}d\eta = dh - \xi(h)\eta \, .
\end{equation}
The vector field \(X_h\) in the local coordinates is expressed as,
\begin{equation}\label{contactfield}
  X_h = \bigg(p_i\frac{\partial h}{\partial p_i}-h\bigg)\frac{\partial}{\partial s} - \bigg(p_i \frac{\partial h}{\partial s}+\frac{\partial h}{\partial q^i}\bigg)\frac{\partial}{\partial p_i} + \bigg(\frac{\partial h}{\partial p_i}\bigg)\frac{\partial}{\partial q^i}.
\end{equation}
It therefore follows that the flow of vector field \(X_h\) is,
\begin{equation}\label{contacteqns}
  \dot{s} =  p_i\frac{\partial h}{\partial p_i} - h; \hspace{3mm} \dot{q}^i = \frac{\partial h}{\partial p_i}; \hspace{3mm} \dot{p}_i = -p_i \frac{\partial h}{\partial s} - \frac{\partial h}{\partial q^i}.
\end{equation}
These equations resemble the Hamilton's equations of motion. Therefore, given a function \(h:\mathcal{M} \rightarrow \R\), the triple \((\mathcal{M},\eta,h)\) is called a contact Hamiltonian system. The function \(h\) is called a contact Hamiltonian function. It is easy to see that the equations of motion [eqns (\ref{contacteqns})] give rise to dissipative dynamics. In fact, it is straightforward to check that,
\begin{equation}\label{doth}
  X_h(h) = -h\xi(h).
\end{equation}
Therefore, \(h\) is not conserved along the flow of \(X_h\). If however \(h=0\), it stays equal to zero along the flow of the corresponding contact vector field. This means that if and only if \(h=0\), it stays zero along the flow of \(X_h\). Therefore, if \(L\) be a Legendre submanifold and \(h|_L=0\) this means that \(L\) is invariant to the flow of \(X_h\). In other words, \(X_h\) is tangent to \(L\) and flow is restricted to \(L\). Since in the thermodynamic sense, the thermodynamic variables are the coordinates on \(\mathcal{M}\) and \(L\) is the space of thermodynamic equilibrium states, this means that a thermodynamic transformation can be interpreted as the flow of some contact vector field generated by a suitable contact Hamiltonian function that vanishes on \(L\) (see \cite{contactBH} for a contact Hamiltonian approach to black hole thermodynamic processes).

\subsection{Gauss-Bonnet-AdS black holes as deformations of the ideal gas}
\renewcommand{\theequation}{\thesubsection.\arabic{equation}}

In the high temperature regime, black holes admit a behavior like that of an ideal gas from standard thermodynamics. In such an "ideal gas" limit in \(d\) dimensions, black holes admit an equation of state of the form,
\begin{equation}\label{idealeqn}
  PV^{1/d-1} =\frac{d-2}{4}\bigg(\frac{\omega_{d-2}}{d-1} \bigg)^{1/(d-1)} T,
\end{equation}
where \(\omega_{d-2}\) is given as,
\begin{equation}
  \omega_{d-2} = \frac{2\pi^{(d-1)/2}}{\Gamma((d-1)/2)}.
\end{equation}

It is possible to deform the high temperature equation of state to obtain that of Gauss-Bonnet black holes in AdS. Recalling the general relation between thermodynamic volume \(V\) and specific volume \(v\) of a black hole in \(d\) dimensions,
\begin{equation}\label{V-v}
  V = \frac{\omega_{d-2}}{d-1}\bigg(\frac{(d-2)v}{4}\bigg)^{d-1},
\end{equation}
it follows that the ideal gas equation for black holes can be written for all values of \(d\) as,
\begin{equation}\label{PvequalsT}
  Pv = T.
\end{equation}

We shall consider for simplicity of calculations, the charged Gauss-Bonnet-AdS case with \(d=5\) for which the equation of state \cite{Cai:2001dz,GB} becomes in terms of the specific volume \(v\) of the black hole,
\begin{equation}\label{GBeqnofstated=5}
  P = \frac{T}{v}\bigg( 1 + \frac{32\alpha}{9 v^2}\bigg) - \frac{2}{3 \pi v^2} + \frac{512q^2}{243 \pi v^6},
\end{equation}
where the black hole horizon is taken to have a spherical topology and \(\alpha = (d-3)(d-4)\alpha_{GB}=2\alpha_{GB}\) with \(\alpha_{GB}\) being the Gauss-Bonnet coupling constant. Here \(v\) is the specific volume related to the horizon radius as \(3v = 4r_+\). The Gauss-Bonnet-AdS black hole can be shown to emerge from the ideal gas limit if we consider a deformation of the ideal gas equation. For that we consider the contact Hamiltonian function \(h:\mathcal{M} \rightarrow \R\) of the following form,

\begin{equation}\label{H-GB}
  h = A^6\Bigg[\frac{128\alpha T}{9A^3} V^{1/4} - \frac{4}{3 \pi A^4}V^{1/2} - \frac{1024q^2}{243\pi}V^{-1/2}\Bigg],
\end{equation}

where \(A\) is a constant given by,
\begin{equation}\label{constantA}
  A = \frac{3}{4}\bigg(\frac{\pi^2}{2}\bigg)^{1/4}.
\end{equation}
It is quite easy to see from the contact Hamiltonian equations of motion [eqns (\ref{contacteqns})] that in the enthalpy representation the thermodynamic variables \(T\) and \(V\) are conserved along the flow of the corresponding contact vector field [eqn (\ref{contactfield})]. On the other hand, the pressure is not conserved and the corresponding evolution expressed in terms of the specific volume is calculated to be,
\begin{equation}\label{dotPidealtoGB}
  \dot{P} = \frac{32\alpha T}{9v^3} - \frac{2}{3\pi v^2} + \frac{512q^2}{243\pi v^6}.
\end{equation}

This means that one may write,
\begin{equation}\label{Pevolve}
  P(\tau) = P_0 + \frac{32\alpha T}{9v^3}\tau - \frac{2}{3\pi v^2}\tau + \frac{512q^2}{243\pi v^6}\tau,
\end{equation}
where \(\tau \in \R\) is a real parameter and \(P(0) = P_0\). It is easy to see that since \(T=T_0\) and \(v=v_0\), from the ideal gas equation one gets \(P_0 = T_0/v_0 = T/v\) and therefore eqn (\ref{Pevolve}) gives,
\begin{equation}\label{deformed}
  P(\tau) = \bigg(1 + \frac{32\alpha }{9v^2}\tau \bigg)\frac{T}{v} - \frac{2}{3\pi v^2}\tau + \frac{512q^2}{243\pi v^6}\tau.
\end{equation}
It should be noted that in this case, \(h \neq 0\) on the Legendre submanifold representing the black hole ideal gas and therefore, the corresponding contact vector field \(X_h\) is not tangent to the Legendre submanifold. In this case one cannot treat \(h\) as a generator of a thermodynamic process as was suggested in sub-section in Appendix A. In fact, in this case \(h\) should be interpreted as a generator of a family of thermodynamic systems\footnote{Equivalently, family of Legendre submanifolds.} i.e. different charged Gauss-Bonnet-AdS black holes for different values of the real parameter \(\tau\).

\smallskip

We have therefore mapped the high temperature ideal gas limit to a family of Gauss-Bonnet-AdS black holes via a deformation induced by a contact Hamiltonian vector field. We remark that this is not just restricted to five dimensions. One may work in any \(d>5\) and perform such deformations.

\end{document}